\documentclass[12pt]{amsart}

\usepackage{graphicx}
\usepackage[font=small,labelfont=bf]{caption} 

\begin{document}

\title[]{Conserved Restricted Solid-on-Solid model}

\author{Anamaria Savu}
\address{Canadian Vigour Center, 4-120 Katz Group Centre for Pharmacy and Health Research, University of Alberta, Edmonton AB, CANADA T6G 2E1}
\email{savu@ualberta.ca}


\date{August 1, 2016}



\maketitle

\vfill

\newpage

\baselineskip24pt

\vspace*{1mm}

\noindent
ABSTRACT. We study a model for the movement of surfaces, namely the conserved, restricted solid-on-solid model. The surface configurations are restricted such that the difference between the heights at adjacent sites is no more than one. In addition the total number of particles is preserved by the dynamics of the model. Mean-field approximations are used to approximate the one-site probability function of the model.

\newpage
\section{CONSERVED RESTRICTED SOLID-ON-SOLID MODEL}
Solid-on-Solid (SOS) models are interacting particle systems developed to understand the evolution of interfaces when overhangs of the interface and bubbles of the bulk phase are neglected. From a theoretical point of view they are Markov chains or processes with interesting properties. Several SOS models are discussed in both mathematical in physical literature.  In this paper we are concerned with a SOS model with Kawasaki dynamics, a dynamics that preserves the total number of particles. The model, known as the conserved restricted SOS (CRSOS) was first mentioned in the physics literature \cite{Sun}. Next we use mean field approximation techniques to approximate the one-site probability function of the model.

We consider a one-dimensional finite lattice with $n$ sites $1, 2, \dots , n$. The lattice is assumed to be periodic. An interface or surface over this lattice is described by a $n$-dimensional vector $h$, where for each $i$, $h_i$ represents the number of particles sitting at site $i$. The interface is restricted such that the height difference between adjacent sites is $|h_{i+1}-h_i| \leq 1$, for all $1 \leq i \leq n$.

All possible configurations of such an interface form a set of $n$-dimensional, integer-valued vectors: 
$$\Omega_{n,K} = \bigg\{ h \in \mathbb{Z}^n_{+} \; \big | \; \sum_{i=1}^n h_i = K, \; \mathrm{and} \; \; |h_{i+1}-h_i| \leq 1 \quad  \forall \; 1 \leq i \leq n \bigg\}.$$ 

The dynamics is described as follows. We randomly pick a site $i$. The top particle at site $i$ leaves the site and moves to one of the sites $i-2$, $i-1$, $i+1$, and $i+2$ provided that the configuration remains restricted after such change. Otherwise the update is cancelled and the process is continued. 

{\it Transitions between site $i$ and sites $i+2$, $i-2$.} A particle can leave site $i$ whenever, prior to its departure, the surface at sites $i-1$, $i$, and $i+1$ has one of the following four departure profiles (Figure 1)
$$(h_i, h_i, h_i), \quad (h_i-1, h_i, h_i), \quad (h_i, h_i, h_i-1), \quad (h_i-1, h_i, h_i-1).$$
And, the particle can land on site $i+2$ whenever, the surface at sites $i+1$, $i+2$, and $i+3$ has one of the following four arrival profiles (Figure 2):
$$(h_{i+2}, h_{i+2}, h_{i+2}), \quad (h_{i+2}, h_{i+2}, h_{i+2}+1), \quad (h_{i+2}+1, h_{i+2}, h_{i+2}), \quad (h_{i+2}+1, h_{i+2}, h_{i+2}+1).$$ Similarly if the landing site is $i-2$.

{\it Transitions between site $i$ and sites $i+1$, $i-1$.} A particle can leave site $i$ and land on site $i+1$ whenever, prior to its departure, the profile at sites $i$, and $i+1$ is $(h_i, h_i-1)$. Similarly if the landing site is $i-1$.

\begin{center}
\includegraphics [scale=0.5]{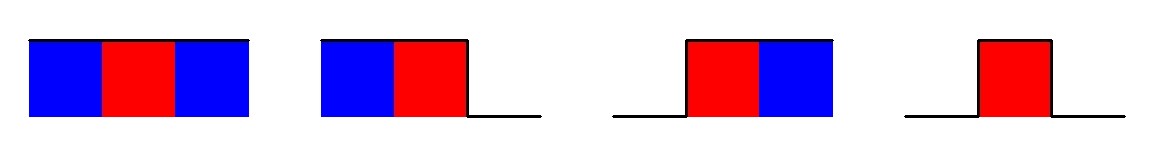}
\captionof{figure}{Departure profiles: the red particle can leave its location}
\end{center}

\begin{center}
	\includegraphics [scale=0.5]{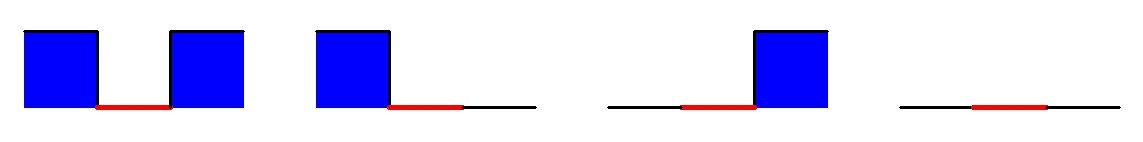}
	\captionof{figure}{Arrival profiles: a particle can arrive at the red site}
\end{center}

The transitions of the process can be categorized into: climb, descend, skip or slide. Each transition is performed with a certain rate as shown below

\begin{center}
	\includegraphics [scale=0.5]{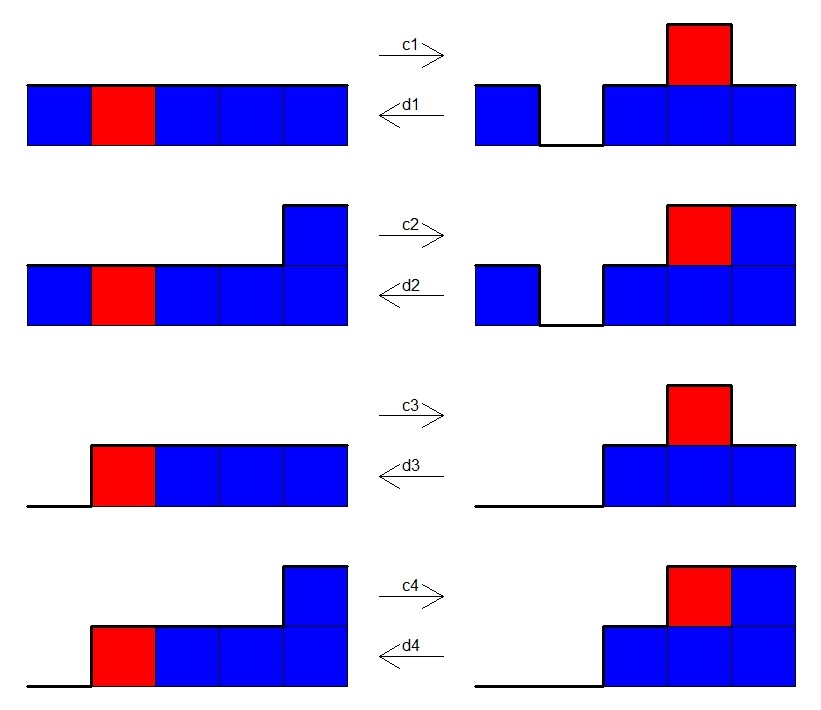}
	\captionof{figure}{Rates for climb and descend transitions }
\end{center}

\begin{center}
	\includegraphics [scale=0.8]{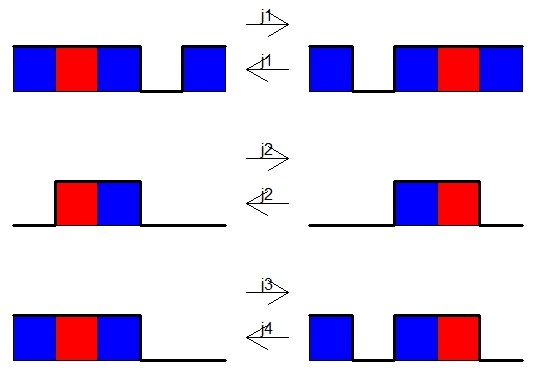}
	\captionof{figure}{Rates for skip transitions }
\end{center}

\begin{center}
	\includegraphics [scale=0.8]{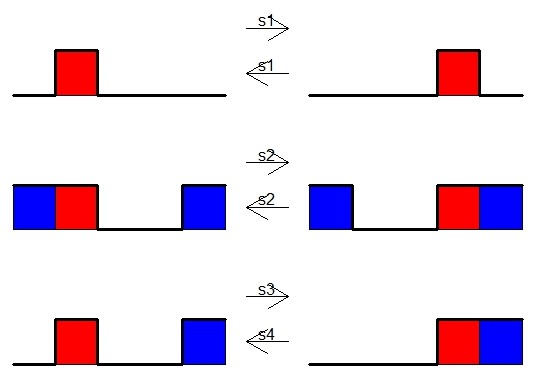}
	\captionof{figure}{Rates for slide transitions on 5 sites}
\end{center}

\begin{center}
	\includegraphics [scale=1]{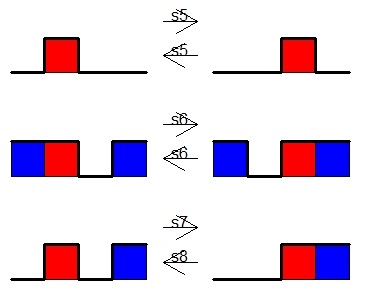}
	\captionof{figure}{Rates for slide transitions on 4 sites}
\end{center}

\section{PROBABILITY FUNCTIONS}

The aim is to characterize the transition probabilities of the process. Specifically, we would like to know the probability that the process is in configuration $h=(h_1, \dots, h_n)$ at time $t$ given that the process started from a certain distribution on height configurations
; probability that we denote by 
$$P_h(t)=P_{(h_1,\dots , h_n)}(t).$$
We write the Kolmogorov forward equation for transition probabilities:
\begin{eqnarray}
\frac{dP_h}{dt}(t) & = & \sum_k P_k(t) a(k,h).
\end{eqnarray}

If the process started from configuration $h^0=(h_1^0, \dots , h_n^0)$ at time $0$ then the initial condition for system (1) is $P_h(0)  =  \delta_{h^0 h}$ for all $h \in \Omega_{n,K}$.
Kolmogorov forward equation is a system of first-order differential equations and involves the rates, $a(k,h)=c(k,h)$, that the process transitions from configuration $k$ to $h$, and $a(h,h)= -\sum_{l \neq h} c(h,l) $. The initial height configuration, $h_0$, can be taken as a flat substrate with $K$ particles distributed uniformly across $n$ sites; $K$ can be taken as a natural number that is a multiple of $n$, equivalently,  $h^0=\big(\frac{K}{n}, \dots , \frac{K}{n}\big)$.

The marginal probability distribution at time $t$ of the height at any given site is defined as
$$P_{h_1} (t) = \sum_{h_2, \dots h_n} P_{(h_1,\dots , h_n)}(t).$$
The time evolution of the probability distribution of one-site height distribution can be derived from Kolmogorov forward equation by summing over $h_2, \dots, h_n$ in both sides of the equation. Because of this summation the terms $P_k(t)c(k,h)$ and $P_h(t)c(h,l)$ cancel each other as long as $k=(k_1, \dots , k_n)$ with $k_1 \neq h_1$ and $l=(l_1, \dots , l_n)$ with $l_1 \neq h_1$. What remains out of the sum (1) are those terms $P_k(t)c(k,h)$ and $P_h(t)c(h,l)$ that correspond to a particle leaving or arriving at site $1$. Therefore the one-site probability function satisfies
\begin{eqnarray}
\frac{dP_{h_1}}{dt}(t) & = & \; \; 2 \times \sum_{h_n,h_2,h_3,h_4} \sum_{\epsilon \pm 1}P_{(h_n, h_1-\epsilon, h_2,h_3+\epsilon, h_4)}(t) \times\\
 & & \quad \quad \quad \quad \quad \quad \times c((h_n, h_1-\epsilon, h_2,h_3+\epsilon, h_4),(h_n,h_1,h_2,h_3,h_4)) \nonumber \\
&  & + 2 \times \sum_{h_n,h_2,h_3} \sum_{\epsilon \pm 1}P_{(h_n, h_1-\epsilon, h_2+\epsilon, h_3)}(t) \times \nonumber \\
& & \quad \quad \quad \quad \quad \quad \times c((h_n, h_1-\epsilon, h_2+\epsilon, h_3),(h_n,h_1,h_2,h_3)) \nonumber \\
 &  & - 2 \times \sum_{h_n,h_2,h_3,h_4} \sum_{\epsilon \pm 1}P_{(h_n, h_1, h_2,h_3, h_4)}(t) \times \nonumber \\
& & \quad \quad \quad \quad \quad \quad \times c((h_n,h_1,h_2,h_3,h_4),(h_n, h_1-\epsilon, h_2,h_3+\epsilon, h_4)) \nonumber \\
&  & - 2 \times \sum_{h_n,h_2,h_3} \sum_{\epsilon \pm 1}P_{(h_n, h_1, h_2, h_3)}(t) \times \nonumber \\
& & \quad \quad \quad \quad \times c((h_n,h_1,h_2,h_3),(h_n, h_1-\epsilon, h_2+\epsilon, h_3)), \nonumber 
\end{eqnarray}

When the process is started in height configuration $h_0$, the initial condition for system (2) is $P_{h_1^0} (0) =1$, and $P_{l}(0)=0$  for all $l \neq h_1^0$. Moreover, if the initial height configuration  is the flat profile, the starting condition for system (2) becomes $P_{\frac{K}{n}} (0)  = 1$ and $P_{l}(0) =  0$ for all $ l \neq \frac{K}{n}$.
The factor $2$ was added above to account for transitions between sites $1$ and $n-1$, and between $1$ and $n$. 

\section{MEAN FIELD APPROXIMATION}
As the ODE system (2) is difficult to solve, we will make a mean field assumption \cite{Bar}; namely, that all space correlations are neglected and $$P_{(h_{i-2}, h_{i-1}, h_i, h_{i+1}, h_{i+2})}(t) =P_{h_{i-2}}(t)P_{h_{i-1}}(t)P_{h_{i}}(t)P_{h_{i+1}}(t)P_{h_{i+2}}(t).$$
Under this assumption the system (2) becomes
\begin{align}
\frac{dP_k}{dt} (t)=2\bigtriangleup \big(& c_1 P_k^5 + c_2 P_k^4 P_{k+1} +c_3 P_{k-1}P^4_k +c_4 P_{k-1}P_k^3P_{k+1} \\
& - d_1 P_{k-1}P_k^3P_{k+1} -d_2 P_{k-1}P_k^2P_{k+1}^2 -d_3 P_{k-1}^2P_k^2P_{k+1} -d_4P_{k-1}^2P_kP_{k+1}^2 \big) \nonumber \\
+4 & \big(c_1 P_k^5 +c_2 P_k^4 P_{k+1}\big)\delta_{k=0}-\big(c_1P_k^5+c_2P_k^4P_{k+1}\big)\delta_{k=1}, \nonumber
\end{align}
with 
$\bigtriangleup P_k =P_{k-1}-2P_k+P_{k+1}$, being the discrete Laplacian 
and $\delta_{k=0}$, $\delta_{k=1}$ being the Dirac functions supported at 0, and 1, respectively. The reductions of (2) to (3) can be explained as follows. Each transition (climb, descend, skip, and slide) contribute with 4 terms. Climb transition with probability $c_1$ will produce: $P_{k+1}^5$ and $-P_k^5$ if the particle climbs from site $1$ to site $3$ and $P_{k-1}^5$ and $-P_k^5$ if the particle climbs from site $3$ to site $1$. In other words, this transition will produce the function $P_k^5$ of the departure configuration $(k,k,k,k,k)$ evaluated with negative sign at the heights of sites $1$, $3$ before departure leading to $-P_k^5$ and $-P_k^5$, and with positive sign at the heights of sites $1$, $3$ after arrival leading to $P_{k+1}^5$ and $P_{k-1}^5$. 

Notice that only the climb and descend transitions contribute towards the form of the system (3).

{\it Time-independent probability function}

Assume that the one-site probability function does not depend on time, and has the form $P_k=(1-\lambda)\lambda^k$ for some $0<\lambda<1$. Then $\lambda$ is the solution of a quadratic equation
\begin{equation}
 (c_2-d_2) \lambda^2-(c_1-d_1+c_4-d_4) \lambda +c_3-d_3 =0.
\end{equation}
Assume $c_2 \neq d_2$. The quadratic equation (4) has exactly one solution in the interval $(0,1)$ if the jumping rates have the property that
$$(c_3-d_3)(c_2-d_2-c_1+d_1-c_4+d_4+c_3-d_3) < 0, $$
and has exactly two solutions in the interval $(0,1)$ if the jumping rates have the properties that 
$$\frac{c_3-d_3}{c_2-d_2} <0 \quad \mathrm{and} \quad  \frac{c_2-d_2-c_1+d_1-c_4+d_4+c_3-d_3}{c_2-d_2} <0 .$$
This corresponds to the case of a smooth surface. The mean height $\overline{h}$ and the square of the interface width $w^2$ are determined as the average and the variance of the distribution $P_k=(1-\lambda)\lambda^k$, that is 
$$\overline{h}=\frac{\lambda}{1-\lambda} \quad \mathrm{and} \quad w^2=\frac{\sqrt{\lambda}}{1-\lambda}.$$
{\it Time-dependent probability function}

The case where the one-site probability function depends on time is solved via another approximation \cite{Gin}. Let $l=\epsilon k $ and $s=\epsilon^{2} t$ be new variables, and define $\widetilde{P}(l,s)=P( \epsilon^{-1} l,\epsilon^{-2} s)=P_{\epsilon^{-1} l}(\epsilon^{-2} s)$. The roles of the new variables are: (1) bring the surface height on a scale from 0 to 1, and (2) speed up the time. We use Taylor expansions in the space variable $l$ to approximate formally $\widetilde{P}$ as a function of $l$ and $s$. Namely for natural numbers $i \geq 1$,

\begin{eqnarray}
\widetilde{P}(l+\epsilon,s) & \approx & \widetilde{P}+\epsilon \frac{\partial \widetilde{P}}{\partial l} + \epsilon^2 \frac{1}{2} \frac{\partial^2 \widetilde{P}}{\partial l^2} \\
\widetilde{P}^i(l+m\epsilon,s) & \approx  &\widetilde{P}^i+\epsilon m i \widetilde{P}^{(i-1)} \frac{\partial \widetilde{P}}{\partial l} + \epsilon^2 \frac{m^2i}{2} \bigg[  \widetilde{P}^{(i-1)} \frac{\partial^2 \widetilde{P}}{\partial l^2} +   (i-1) \widetilde{P}^{(i-2)} \bigg(\frac{\partial \widetilde{P}}{\partial l}\bigg)^2   \bigg] \nonumber
\end{eqnarray}
with the expansions in the right side being evaluated at $(l,s)$. 
Then Kolmogorov forward equation is approximated by
\begin{equation}
 \frac{\partial \widetilde{P}}{\partial s} = A \times \bigg[ 5 \widetilde{P}^{4}(l,s) \frac{\partial^2 \widetilde{P}}{\partial l^2}  +  20\widetilde{P}^{3}(l,s) \bigg( \frac{\partial \widetilde{P}}{\partial l}\bigg)^2   \bigg].
 \end{equation}
for some constant $A$ that depends on the jumping rates.
Next we search for a solution of (6) of the form, \cite{Fam}, 
$$ \widetilde{P}(l,s) = s^{-\gamma} f ( l \, s^{-\gamma}).$$
Then 
\begin{eqnarray}
\frac{\partial \widetilde{P}}{\partial s}(l,s) & = & -\gamma \, s^{-\gamma -1}f( l \, s^{-\gamma})  -\gamma \, l \,  s^{-2\gamma -1}f'( l \, s^{-\gamma})  \\
\frac{\partial \widetilde{P}}{\partial l}(l,s) & =  & s^{-2\gamma}f'( l \, s^{-\gamma})   \nonumber \\
\frac{\partial^2 \widetilde{P}}{\partial l^2}(l,s) & =  & s^{-3\gamma}f^{''}( l \, s^{-\gamma}).  \nonumber
\end{eqnarray}
After rewriting (6) using (7) we have that for some constant $A$
\begin{align}
(-\gamma \, s^{-\gamma -1}) & \times \big[f( l \, s^{-\gamma})  +l \,  s^{-\gamma }f'( l \, s^{-\gamma})\big]  \\
& = As^{-7\gamma} \times \big[5 f^4( l \, s^{-\gamma})f^{''}( l \, s^{-\gamma})  +20  f^3( l \, s^{-\gamma})(f'( l \, s^{-\gamma}))^2\big]. \nonumber
\end{align}
If we choose $\gamma=1/6$, such that $-\gamma-1=-7\gamma$, and $x= l \, s^{-\gamma}$ we find that (8) becomes 
$$\gamma \big[f(x)+xf'(x)\big]+A\big[5f^4(x)f^{''}(x)+20 f^3(x)(f'(x))^2\big]=0,$$
$$ \gamma \,\big[ xf(x)\big]'+A \big[f^5(x)\big]^{''}=0,$$
$$ \gamma \, xf(x) +A \big[f^5(x)\big]^{'}=C_0.$$
If we pick the constant $C_0$ to be equal to $0$ then the above equation is separable and can be shown to have the following solution
$$ f(x)=\sqrt[4]{C_1-\frac{x^2}{15A}},$$
that depends on the constant $C_1$. Hence
$$ \widetilde{P}(l,s) = \sqrt[4]{\frac{C_1}{s^{1/3}}-\frac{1}{15A} \times \frac{l^2}{s^{2/3}}} . $$
Since $l=\epsilon k$ and $s=\epsilon^2 t$ we find that the time-dependent one-site probability function is
$$ P_k(t)=\sqrt[4]{\frac{C_1}{\epsilon^{1/2}}\times\frac{1}{t^{1/3}}-\frac{\epsilon}{15A} \times\frac{k^2}{t^{2/3}}} $$

This corresponds to a rough surface. The average and variance of the above distribution can be calculated to find that the mean height $\overline{h}\sim t^{1/12}$ and the width $w^2 \sim t^{1/4}$.

\end{document}